\documentstyle[12pt]{article}
\begin{document}
\parskip 10pt plus 1pt
\title{THE MEANING OF MAXIMAL SYMMETRY IN  PRESENCE OF TORSION}
\date {}
\maketitle
\centerline{\it  Debashis Gangopadhyay\footnote{S.N.Bose National Centre
For Basic Sciences,JD Block, Sector-III, Salt Lake, Calcutta-700091, INDIA.
e-mail :debashis@bose.ernet.in}
and Soumitra Sengupta\footnote {Department of Physics, Jadavpur University,
Calcutta-700032, INDIA} }
\baselineskip=20pt

\newpage
\begin{abstract}
We show that the usual physical meaning of maximal symmetry can be made to
remain unaltered even if torsion is present. All that is required is that
the torsion fields satisfy some mutually consistent constraints. We also
give an explicit realisation of such a scenario by determining the torsion
fields , the metric and the associated Killing vectors.

PACS NO: 04.20-q ; 11.17+y
\end{abstract}

\newpage
In $N$ dimensions, a metric that admits the maximum number $N(N + 1)/2$ of 
Killing vectors is said to be maximally symmetric.A maximally symmetric space 
is homogeneous and isotropic about all points [1].Such spaces are
of natural interest in the general theory of relativity as they correspond to 
spaces of globally constant curvature which in turn is related to the concepts 
of homogeneity and isotropy.The requirement of isotropy and homogeneity leads
to maximally symmetric metrics in the context of standard cosmologies -- the
most well known example being the Robertson-Walker cosmology.However, in the 
presence of torsion there is a drastic change in the scenario and one needs 
to {\it redefine} maximal symmetry itself.

In this paper we propose a  way to do this such that {\it the usual physical
meaning of maximal symmetry remains the same.} The only requirement is that
the torsion fields satisfy some mutually consistent constraints. We also
give an example of a toy model where our ideas can be realised by 
determining the torsion fields, the metric and the Killing vectors.

The presence of torsion implies that the affine connections 
$\bar{\Gamma}^{\alpha}_{\mu\nu}$ are asymmetric and contain an antisymmetric
part $ H ^{\alpha}_{\mu\nu}$ in addition to the usual symmetric term 
$\Gamma^{\alpha}_{\mu\nu}$  [2,3]:
$$\bar{\Gamma}^{\alpha}_{\mu\nu} = \Gamma^{\alpha}_{\mu\nu} + H^{\alpha}_{\mu\nu}\eqno(1)$$
The notation has been chosen keeping string theory in mind where the 
antisymmetric third rank tensor $H_{\alpha\mu\nu} = \partial_{(\alpha} B_{\mu\nu)}$
is identified with the background torsion and $B_{\mu\nu}$ is the second rank
antisymmetric tensor field mentioned above.Note that $H^{\alpha}_{\mu\nu}$ are
completely arbitrary to start with.

Defining covariant derivatives with respect to $\bar{\Gamma}^{\alpha}_{\mu\nu}$
we have for a vector field $V_{\mu}$ :
$$V_{\mu;\nu;\beta} - V_{\mu;\beta;\nu} = -\bar{R}^{\lambda}_{\mu\nu\beta}
V_{\lambda} + 2 H^{\alpha}_{\beta\nu} V_{\mu;\alpha} \eqno(2)$$
where 
$$\bar{R}^{\lambda}_{\mu\nu\beta} = R^{\lambda}_{\mu\nu\beta} + 
\tilde{R}^{\lambda}_{\mu\nu\beta} \eqno(3a)$$
$$R^{\lambda}_{\mu\nu\beta} = \Gamma^{\lambda}_{\mu\nu, \beta} -
\Gamma^{\lambda}_{\mu\beta, \nu} + 
\Gamma^{\alpha}_{\mu\nu}\Gamma^{\lambda}_{\alpha\beta} -
\Gamma^{\alpha}_{\mu\beta}\Gamma^{\lambda}_{\alpha\nu} \eqno(3b)$$
$$\tilde{R}^{\lambda}_{\mu\nu\beta} = H^{\lambda}_{\mu\nu, \beta} -
H^{\lambda}_{\mu\beta, \nu} +
H^{\alpha}_{\mu\nu}H^{\lambda}_{\alpha\beta} - 
H^{\alpha}_{\mu\beta}H^{\lambda}_{\alpha\nu}$$
$$\phantom{\tilde{R}^{\lambda}_{\mu\nu\beta}} +
\Gamma^{\alpha}_{\mu\nu}H^{\lambda}_{\alpha\beta} - 
H^{\alpha}_{\mu\beta}\Gamma^{\lambda}_{\alpha\nu} +
H^{\alpha}_{\mu\nu}\Gamma^{\lambda}_{\alpha\beta} -
\Gamma^{\alpha}_{\mu\beta}H^{\lambda}_{\alpha\nu} \eqno(3c) $$
The generalised curvature $\bar{R}^{\lambda}_{\mu\nu\beta}$ does not have the
usual symmetry (antisymmetry) properties.Note that the last term on the 
right hand side $(2)$ is obviously a tensor.Hence 
$\bar{R}^{\lambda}_{\mu\nu\beta}$ is also a tensor.

Now, the determination of all infinitesimal isometries of a metric is 
equivalent to determining all Killing vectors $\xi_{\mu}$ of the metric.A
Killing vector is defined through the Killing condition:
$$\xi_{\mu\enskip ;\enskip\nu} + \xi_{\nu\enskip ;\enskip\mu} = 0\eqno(4)$$
and it is easily verified that this condition is preserved also in the 
presence of the torsion $H^{\alpha}_{\mu\nu}$.

Equation $(2)$ for a Killing vector hence takes the form:
$$\xi_{\mu;\nu;\beta} - \xi_{\mu;\beta;\nu} = -
\bar{R}^{\lambda}_{\mu\nu\beta} \xi_{\lambda} +
2 H^{\alpha}_{\beta\nu}  \xi_{\mu;\alpha}\eqno(5)$$
The $H^{\alpha}_{\beta\nu}$ are arbitrary and so we may choose them to be 
such that 
$$H^{\alpha}_{\beta\nu}  \xi_{\mu ;\alpha} = 0 \eqno(6)$$
This is a constraint on the $H^{\alpha}_{\beta\nu}$ and not the Killing
vectors  $\xi_{\mu}$.{\it We shall shortly see that this constraint is 
essential for the existence of maximal symmetry even in the presence of 
torsion.} The commutator of two covariant derivatives of a
Killing vector thus becomes:
$$\xi_{\mu;\nu;\beta} - \xi_{\mu;\beta;\nu} = -
\bar{R}^{\lambda}_{\mu\nu\beta} \xi_{\lambda}\eqno(7)$$

We now impose the cyclic sum rule on $\bar{R}^{\lambda}_{\mu\nu\beta}$:
$$\bar{R}^{\lambda}_{\mu\nu\beta} + \bar{R}^{\lambda}_{\nu\beta\mu} + 
\bar{R}^{\lambda}_{\beta\mu\nu} = 0 \eqno(8a)$$
As $R^{\lambda}_{\mu\nu\beta} +R^{\lambda}_{\nu\beta\mu} +
R^{\lambda}_{\beta\mu\nu} = 0$  the constraint $(8a)$ implies 
$$\tilde{R}^{\lambda}_{\mu\nu\beta} + \tilde{R}^{\lambda}_{\nu\beta\mu} +
\tilde{R}^{\lambda}_{\beta\mu\nu} = 0 $$
and this reduces to 
$$ H^{\lambda}_{\mu\nu,\beta} + 
H^{\alpha}_{\mu\nu}\bar{\Gamma}^{\lambda}_{\alpha\beta} +
H^{\lambda}_{\nu\beta,\mu} + 
H^{\alpha}_{\nu\beta}\bar{\Gamma}^{\lambda}_{\alpha\mu} +
H^{\alpha}_{\beta\mu,\nu} +
H^{\alpha}_{\beta\mu}\bar{\Gamma}^{\lambda}_{\alpha\nu} = 0\eqno(8b) $$

Hence adding $(7)$ and its two cyclic permutations and using $(4)$, the 
relation $(7)$ can be recast into 
$$\xi_{\mu;\nu;\beta} = - 
\bar{R}^{\lambda}_{\beta\nu\mu}\xi_{\lambda}\eqno(9)$$

Therefore given $\xi_{\lambda}$ and $\xi_{\lambda;\nu}$ at some point $X$ ,
we can determine the second derivatives of $\xi_{\lambda}$$(x)$ at $X$ from
$(9)$.Then following the usual arguments [1], any Killing vector 
$\xi^{n}_{\mu}$$(x)$ of the metric $g_{\mu\nu}$$(x)$ can be expressed as 
$$\xi^{n}_{\mu}(x) = A^{\lambda}_{\mu}(x\enskip ;\enskip X)
\xi^{n}_{\lambda}(X) + C^{\lambda\nu}_{\mu} (x\enskip;\enskip X)
\xi^{n}_{\lambda ;\nu}(X)\eqno(10)$$
where $A^{\lambda}_{\mu}$ and $C^{\lambda\nu}_{\mu}$ are functions that 
depend on the metric, {\it torsion} and $X$ but not on the initial values 
$\xi_{\lambda}$$(X)$ and $\xi_{\lambda;\nu}$$(X)$, and hence are the same 
for all Killing vectors.Also note that the torsion fields present in 
$A^{\lambda}_{\mu}$$(x ; X)$ and  $C^{\lambda\nu}_{\mu}$$(x ; X)$ obey the 
constraint $(6)$.A set of Killing vectors $\xi^{n}_{\mu}$$(x)$  is said to 
be independent if they do not satisfy any relation of the form 
$\Sigma_{n} d_{n} \xi^{n}_{\mu}$$(x)$ $= 0$,with constant coefficients 
$d_{n}$.It therefore follows that there can be at most $N(N+1)/2$ 
independent Killing vectors in $N$ dimensions,{\it even in the presence of 
torsion provided the torsion fields satisfy the constraints (6) and (8b).
Therefore, one can generalise the concept of maximal symmetry to the case where
torsion is present , provided the torsion fields satisfy the constraints 
embodied in the equations (6) and (8b).} This is the principal result of this
paper.

Now, $\bar{R}_{\lambda\mu\nu\beta}$ is antisymmetric in the indices 
$(\lambda, \mu)$  and $(\nu, \beta)$.This follows from the fact that 
$R_{\lambda\mu\nu\beta}$  and $\tilde{R}_{\lambda\mu\nu\beta}$ both have 
these properties.This can be verified through an elaborate but 
straightforward calculation.Then proceeding as in Ref.[1],we have 
(using -$\bar{R}^{\alpha}_{\mu\nu\alpha} = \bar{R}_{\mu\nu}$ etc.)
$$(N-1) \bar{R}_{\lambda\mu\nu\beta} = \bar{R}_{\beta\mu}g_{\lambda\nu} -
\bar{R}_{\nu\mu}g_{\lambda\beta}$$
i.e.$$(N-1) R_{\lambda\mu\nu\beta} + (N-1) \tilde{R}_{\lambda\mu\nu\beta}$$ 
$$= R_{\beta\mu} g_{\lambda\nu} - R_{\nu\mu} g_{\lambda\beta} +
\tilde{R}_{\beta\mu} g_{\lambda\nu} - 
\tilde{R}_{\nu\mu} g_{\lambda\beta}\eqno(11)$$
where $N$ is the number of dimensions.
$R_{\lambda\mu\nu\beta}$, $R_{\beta\mu}$ are functions of the symmetric affine
connections $\Gamma$ only, whereas $\tilde{R}_{\lambda\mu\nu\beta}$,
$\tilde{R}_{\beta\mu}$ are functions of both $\Gamma$ and $H$. Broadly,the
solution space of equation (11) consists of (a) solutions with $H$ determined
by $\Gamma$ or vice versa (b) solutions where $H$ and $\Gamma$ are independent
of each other.{\it All these solutions lead to maximally symmetric spaces 
even in the presence of torsion.}.

We shall now illustrate that the smaller subspace (b) of these solutions
enables one {\it to cast the definition of 
maximal symmetry in the presence of torsion in an exactly
analogous way to that in the absence of torsion.}
A particular set of such solutions of  $(11)$ can be obtained 
by  equating corresponding terms on both sides to get: 
$$(N-1) R_{\lambda\mu\nu\beta} = R_{\beta\mu} g_{\lambda\nu} -
R_{\nu\mu} g_{\lambda\beta}\eqno(12a)$$
$$(N-1) \tilde{R}_{\lambda\mu\nu\beta} = \tilde{R}_{\beta\mu} g_{\lambda\nu} -
\tilde{R}_{\nu\mu} g_{\lambda\beta}\eqno(12b)$$

The above two equations lead to
$$R_{\lambda\mu\nu\beta} = R^{\alpha}_{\alpha}(g_{\lambda\nu} g_{\mu\beta} -
g_{\lambda\beta} g_{\nu\mu})/ N(N - 1)\eqno(13a)$$
and 
$$\tilde{R}_{\lambda\mu\nu\beta} = \tilde{R}^{\alpha}_{\alpha}
(g_{\lambda\nu} g_{\mu\beta} - g_{\lambda\beta} g_{\nu\mu}/ N(N -1)\eqno(13b)$$

It is appropriate to note here that 
$$\tilde{R}^{\alpha}_{\alpha\nu\beta} = 0 \eqno(14a)$$
$$\tilde{R}^{\alpha}_{\mu\nu\alpha} = H^{\alpha}_{\mu\nu,\alpha} -
H^{\alpha}_{\mu\gamma} H^{\gamma}_{\alpha\nu} + 
H^{\alpha}_{\mu\nu} \Gamma^{\gamma}_{\alpha\gamma} -
H^{\alpha}_{\gamma\nu} \Gamma^{\gamma}_{\mu\alpha} -
H^{\alpha}_{\mu\gamma} \Gamma^{\gamma}_{\alpha\nu} \eqno(14b)$$

Consider the constraint $(8b)$ with $\beta=\lambda$.This gives :  
$$H^{\alpha}_{\mu\nu,\alpha} + 
H^{\alpha}_{\mu\nu} \Gamma^{\gamma}_{\alpha\gamma} +
H^{\alpha}_{\nu\gamma} \Gamma^{\gamma}_{\alpha\mu} -
H^{\alpha}_{\mu\gamma}  \Gamma^{\gamma}_{\alpha\nu} = 0\eqno(15)$$
It is straightforward to verify that this implies
$\tilde{R}_{\mu\nu} = \tilde{R}_{\nu\mu}$. Hence $(8b)$ implies that 
$\tilde{R}_{\mu\nu}$ is symmetric.

Under these circumstances we have 
$$R_{\mu\nu} = (1/N) g_{\mu\nu}R^{\alpha}_{\alpha}\eqno(16a)$$
$$\tilde{R}_{\mu\nu} = (1/N) g_{\mu\nu}\tilde{R}^{\alpha}_{\alpha}
= H^{\alpha}_{\mu\beta} H^{\beta}_{\alpha\nu}\eqno(16b)$$

Now using arguments similar to those given in Ref.[1] for the Bianchi 
identities we can conclude that in the presence of torsion fields 
satisfying constraints discussed before :
$$\bar{R}_{\lambda\mu\nu\beta} = 
R_{\lambda\mu\nu\beta} + \tilde{R}_{\lambda\mu\nu\beta} =
(K+ \tilde{K})(g_{\lambda\nu}g_{\mu\beta} - g_{\lambda\beta}g_{\mu\nu}) =
\bar{K}(g_{\lambda\nu}g_{\mu\beta} - g_{\lambda\beta}g_{\mu\nu})\eqno(17)$$
where
$$R^{\alpha}_{\alpha} = constant = K N (1 - N)\eqno(18a)$$
$$\tilde{R}^{\alpha}_{\alpha} = H^{\mu\lambda}_{\beta} H^{\beta}_{\lambda\mu}=
constant = \tilde{K} N (1 - N)\eqno(18b)$$
$$\bar{K} = K + \tilde{K} = constant\eqno(18c)$$
(In deriving the above results from  the Bianchi identities we have used 
the fact that for a flat metric the curvature constant $K=0$.Hence demanding
$\bar{K}=0$ for a (globally) zero curvature space means that $\tilde{K}=0$
which in turn means that the torsion must vanish.)

We now give a simple example of a torsion field which satisfies $(8b)$ and$(18b)$
and is also consistent with $(6)$.First note that {\it any } non vanishing
torsion is always consistent with $(6)$ because
$$H^{\alpha}_{\beta\nu}  \xi_{\mu ;\alpha} = 0 $$
implies  
$$H^{\alpha}_{\beta\nu}  \xi_{\alpha ;\mu} = 0 \eqno(6b)$$
through the Killing condition (4).Adding (6) and (6b) gives 
$$H^{\alpha}_{\beta\nu}(\xi_{\mu ;\alpha} + \xi_{\alpha ; \mu}) = 0 $$
Hence {\it any non-zero torsion} is consistent with $(6)$ and $(6b)$.
The torsion is an antisymmetic third rank tensor obtained from a second rank
antisymmetric tensor $B_{\mu \nu}$ as follows:
$$H^{\alpha}_{\mu\nu}= \partial^{\alpha}B_{\mu \nu}
+ \partial_{\mu}B_{\nu}^{\alpha} + \partial_{\nu}B^{\alpha}_{\mu}$$

Consider dimension $D=3$ and a general form for the metric as 
$$ds^2 = -f_{0}(r) dt^2 + f(r) dr^2 + f_{1}(r) (dx^1)^2 \eqno(19)$$
So the metric components are 
$$g_{00}=-f_{0}(r), g_{rr}=f(r),  g_{11}= f_{1}(r)$$
i.e. the metric components are functions of $r$ only.Further assume that 
all fields, including $B_{\mu \nu}$ , depend only on $r$.Then the torsion is 
just $H^{r}_{01}$.

It is straightforward to verify that with our chosen metric the only
non-zero components of $\Gamma^{\alpha}_{\mu \nu}$ are 
$\Gamma^{r}_{00}$, $\Gamma^{r}_{11}$, $\Gamma^{r}_{rr}$, 
$\Gamma^{0}_{0r}$ and  $\Gamma^{1}_{1r}$.

Now $(8b)$ with  $\lambda=\beta$ gives $(15)$ which in the case under 
consideration reduces to (using the antisymmetry of $H$)
$$H^{r}_{01,r} + H^{r}_{01} \Gamma^{r}_{rr} + H^{0}_{1r} \Gamma^{r}_{00} 
- H^{1}_{0r} \Gamma^{r}_{11} = 0 $$ 

We may write
$$ H^{0}_{1r} = g^{00}g_{rr} H^{r}_{01},
\enskip  H^{1}_{tr} =-g^{11}g_{rr} H^{r}_{01}$$
Therefore $(15)$ becomes
$$H^{r}_{01,r} + H^{r}_{01}\enskip [\Gamma^{r}_{rr}+g^{00}g_{rr}\Gamma^{r}_{00}
+g^{11}g_{rr}\Gamma^{r}_{11}] = 0   \eqno(20)$$
Using the values 
$$\Gamma^{r}_{rr}=\partial_{r}(ln f^{1/2}),
\enskip \Gamma^{r}_{00}= (1/2)(1/f)\partial_{r} f_{0},
\enskip \Gamma^{r}_{11}=-(1/2)(1/f)\partial_{r} f_{1} $$
yields the equation 
$$H^{r}_{01,r} + H^{r}_{01} \partial_{r}[ln(f/(f_{0}f_{1}))^{1/2}] = 0\eqno(21)$$
whose  solution  is 
$$H^{r}_{01} =  [f_{0} f_{1}/ f ]^{1/2} \eqno(22)$$
(Note that the torsion can be taken proportional to the completely antisymmetric
$\epsilon$ tensor in three dimensions as follows : 
$H_{r01} = g_{rr} H^{r}_{01}$ ,
and so $H_{r01}$  may be written as 
$H_{r01} = [ f f_{0} f_{r} ] ^ {1/2} \epsilon_{r01}$ and this can be
further integrated to give the "magnetic field" as 
$B_{01} = \epsilon_{01} \int dr [f f_{0} f_{r}]^{1/2}$ , etc.)

It is immediately verified that 
$$H^{r}_{01}H^{01}_{r} = - 1 \enskip (i.e.a\enskip constant) $$
thereby satisfying the constraint $(18b)$.

If  $\xi$ denote the  the Killing vectors then the Killing equations are
$$\xi^{r} \partial_{r}f_{0} + 2 f_{0} \partial_{t}\xi^{t} = 0\eqno(23a)$$
$$\xi^{r} \partial_{r}f + 2 f \partial_{r}\xi^{r} = 0\eqno(23b)$$
$$\xi^{r} \partial_{r}f_{1} + 2 f_{1} \partial_{1}\xi^{1} = 0\eqno(23c)$$
$$f_{1}\partial_{t}\xi^{1}-  f_{0} \partial_{1}\xi^{t} = 0\eqno(23d)$$
$$f\partial_{t}\xi^{r}-  f_{0} \partial_{r}\xi^{t} = 0\eqno(23e)$$
$$f_{1}\partial_{r}\xi^{1}+ f \partial_{1}\xi^{r} = 0\eqno(23f)$$

The solutions to these equations are 
$$f_{0} = Exp[-2\alpha\int dr f^{1/2}]\eqno(24a)$$
$$f_{1} = Exp[-2\gamma\int dr f^{1/2}]\eqno(24b)$$
$$\xi^{r} = 1/f^{1/2}\eqno(24c)$$
$$\xi^{t} = \alpha t + \eta (x^{1}) + \beta \eqno(24d)$$
$$\xi^{1} = \gamma x^{1} + \psi(t) + \delta \eqno(24e)$$
where $\alpha , \beta, \gamma, \delta $ are constants and $\eta(x^{1})$
and $\psi(t)$ are functions of $x^{1}$ only and $t$ only respectively.

We therefore have an explicit example of a scenario where our generalised
definition of maximal symmetry works. Note that the metric coefficients
can be any well behaved functions of $r$ and the torsion is related to these
coefiicients (equation $(22)$).This is as it should be and the rather general
nature of these solutions (i.e.not limited to any unique function of $r$)
imply that our results are neither accidental nor ad-hoc.

Therefore, in the presence of torsion the criteria of maximal symmetry has 
been generalised through the equations $(13b)$, $(17)$ and $(18)$.The physical
meaning is still that of a globally constant curvature (which now also has a 
contribution coming from the torsion).The torsion fields are, however, 
subject to the constraints embodied in equations $(6)$, $(8b)$ and $(18b)$ 
and these constraints are mutually consistent.Hence our usual concepts of 
the  homogeneity and isotropy of space have a more generalised footing in 
the presence of torsion.The thing to note is that the existence of torsion 
does not necessarily jeopardise the prevalent concepts of an isotropic and 
homogeneous spacetime.All that is required is that the torsion fields obey 
certain mutually consistent constraints.These constraints, in some sense, 
ensure that the usual physical meaning of an isotropic and homogeneous 
spacetime remains intact.Hence, in the context of the General Theory of 
Relativity,the present work generalises the concept of maximal symmetry
of spacetimes.

It is also worth mentioning here  the relevance of the present work in the 
context of string theory.The low energy string effective action posseses
for time dependent metric $G_{\mu\nu} (\mu,\nu = 1,2,....d)$, torsion 
$B_{\mu\nu}$ and dilaton $\phi$ background fields a full continuous 
$O(d,d)$ symmetry under which "cosmological" solutions of the equations of 
motion are transformed into other inequivalent solutions [4].A solution
with zero torsion is necessarily connected to a solution with
non-zero torsion.Therefore, in the framework of the 
generalised maximal symmetry discussed here it is worth studying whether this 
generalised maximal symmetry is preserved under the $O(d)\otimes O(d)$
transformation.The results of this investigation will be reported elsewhere.

Standard Cosmology largely supports the homogeneity and isotropy of the 
universe.Our results show that the presence of torsion does not necessarily
jeopardise this fact.Finally, the most important consequence of this work
is in the realm of cosmologies with torsion. It would be interesting to see
whether the generalised maximally symmetric solutions can become suitable 
candidates for already existing cosmological observations.

We thank A.K.Raychaudhuri for illuminating discussions.

\end{document}